\documentclass[prl,twocolumn,superscriptaddress,prl]{revtex4-1}

\usepackage{amsmath}
\usepackage{amssymb}
\usepackage{xspace}
\usepackage{graphicx}
\usepackage{grffile}
\usepackage{nicefrac}
\usepackage{color}

\begin{document}

\title{Thermopower enhancement from engineering the Na$_{0.7}$CoO$_2$ interacting
fermiology via Fe doping}

\author{Raphael Richter}
\affiliation{I. Institut f\"ur Theoretische Physik, Universit\"at Hamburg, 
20355 Hamburg, Germany}
\author{Denitsa Shopova}
\affiliation{Chemische Materialsynthese, 
Institut f\"ur Materialwissenschaft, Universit\"at
Stuttgart, 70569 Stuttgart, Germany}
\author{Wenjie Xie}
\affiliation{Chemische Materialsynthese, 
Institut f\"ur Materialwissenschaft, Universit\"at
Stuttgart, 70569 Stuttgart, Germany}
\author{Anke Weidenkaff}
\affiliation{Chemische Materialsynthese, 
Institut f\"ur Materialwissenschaft, Universit\"at
Stuttgart, 70569 Stuttgart, Germany}
\author{Frank Lechermann}
\affiliation{I. Institut f\"ur Theoretische Physik, Universit\"at Hamburg, 
20355 Hamburg, Germany}
\affiliation{Institut f\"ur Keramische Hochleistungswerkstoffe, Technische
Universit\"at Hamburg-Harburg, 21073 Hamburg, Germany}



\begin{abstract}
The sodium cobaltate system Na$_{x}$CoO$_2$ is a prominent representant of strongly
correlated materials with promising thermoelectric response. In a combined theoretical 
and experimental study we show that by doping the Co site of the compound at $x$=0.7 
with iron, a further increase of the Seebeck coefficient is achieved. The Fe defects 
give rise to effective hole doping in the high-thermopower region of larger sodium 
content $x$. Originally filled hole pockets in the angular-resolved spectral function 
of the material shift to low energy when introducing Fe, leading to a 
multi-sheet interacting Fermi surface. Because of the higher sensitivity of correlated
materials to doping, introducing adequate substitutional defects is thus a promising 
route to manipulate their thermopower.
\end{abstract}

\maketitle

\paragraph{Introduction.}
Selected compounds subject to strong electronic correlations display a remarkable
thermoelectric response. Layered cobaltates composed of stacked triangular CoO$_2$ 
sheets are of particular interest because of their large doping range. High thermopower
above 75$\mu$V/K has originally been detected in the Na$_{x}$CoO$_2$ system at 
larger sodium doping $x\sim 0.7$~\cite{ter97,mot01,kau09,lee06}. Even higher Seebeck 
coefficients and increased figure of merit have been measured for misfit cobaltates 
(see e.g.~\cite{heb13} for a recent review). Due to the simpler crystal structure, 
the small-unit-cell sodium-compound system remains of key interest in 
view of the essential electronic-structure effects underlying the pronounced
thermoelectric response.

The high-thermopower region is associated with clear signatures of strong electronic 
correlations. Charge disproportionation~\cite{muk05,lan08}, change of Pauli-like magnetic 
susceptibility to Curie-Weiss-like behavior~\cite{foo04} and eventual onset of in-plane 
ferromagnetic (FM) order~\cite{mot03,boo04,sak04,bay05,hel06,shu07,lan08,schu08} 
at $x$=0.75 are observed. Moreover the region displays unique low-energy 
excitations~\cite{wil15}.

Several theoretical works addressed the thermoelectricity of 
Na$_{x}$CoO$_2$~\cite{kos01,xia07,ham07,kur07,pet07,wis10,san12}, ranging from  
applications of Heikes formula, Boltzmann-equation approaches as well as 
Kubo-formula-oriented modelings. Correlation effects described beyond 
conventional density functional theory (DFT) indeed increase the thermopower. 
Depending on $x$, the oxidation state of cobalt nominally reads 
Co$^{(4-x)+}$($3d^{5+x}$). A strong cubic crystal field establishes a Co 
($t_{2g}$,$e_g$) low-spin state, and $x$ controls the filling of the localized 
$t_{2g}$ manifold. The additional trigonal crystal field splits $t_{2g}$ into 
$a_{1g}$ and $e_g'$ levels. But the measured Fermi surface (FS) shows only a single 
distinct hole-like $a_{1g}$-dominated sheet centered around the $\Gamma$ 
point~\cite{has04,gec07}. Hole pockets of mainly $e_g'$ kind are surpressed by 
correlation effects~\cite{wan08,boe14}.

In this work we show that there is the possibility to engineer the interacting 
electronic structure of Na$_{x}$CoO$_2$ in view of an extra increase of the
thermoelectric response. Substitutional doping of the Co site with Fe is for 
$x$$\sim$0.7 effective in shifting the $e_g'$ hole pockets to the Fermi level 
$\varepsilon_{\rm F}$. The hole doping with iron also enforces the correlation 
strength. But different from the small-$x$ region where the thermopower is found to 
be rather weak, for Na$_{0.7}$Co$_{1-y}$Fe$_y$O$_2$ the Seebeck coefficient is 
further enhanced. This finding paves the road for future design of 
thermoelectric transport in correlated materials.

\paragraph{Theoretical approach}.
An advanced charge self-consistent DFT + dynamical mean-field theory (DMFT) 
scheme~\cite{gri12} is utilized. The method is based on an efficient combination of 
the mixed-basis pseudopotential framework~\cite{lou79} with continuous-time quantum 
Monte-Carlo in the hybridization-expansion representation~\cite{rub05,wer06,boe11,par15} 
for the DMFT impurity problem. The correlated subspace consists of 
projected~\cite{ama08,ani05} Co/Fe $3d(t_{2g})$ orbitals, i.e. a threefold 
many-body treatment holds in the single-site DMFT part. The generic 
multi-orbital Coulomb interaction is chosen in Slater-Kanamori parametrization
with Hubbard $U$=5eV and Hund's exchange $J_{\rm H}$=0.7eV. A double-counting 
correction of the fully-localized form~\cite{ani93} is used. We construct 
Na as well as Co pseudopotentials with fractional nuclear charge to cope
with the doping scenario in a virtual-crystal approximation (VCA). 
Based on a primitive hexagonal cell for one formula unit Na$_x$CoO$_2$, the 
fractional-charged Na is in the so-called 'Na2' position, i.e. aside from the
transition-metal position below. No bilayer splitting is included. The calculations 
are readily extendable to more complex unit cells and geometries, however the 
present approach suits the purpose of rendering general qualitative statements.

For the investigation of transport, the Kubo formalism based on the 
correlation functions 
\begin{align}
K_n=\sum_{\bf k}\int d\omega&\,\beta^n(\omega-\mu)^n
\left(-\frac{\partial f_\mu}{\partial\omega}\right)\notag\times\\
&\times\operatorname{Tr}\left[{\bf v}({\bf k})\,A({\bf k},\omega)\,
{\bf v}({\bf k})\,A({\bf k},\omega)\right]\;, \label{eqn:An}
\end{align}
is used in the DMFT context~\cite{oud06,wis10,den13,boe14}. Here $\beta=1/T$ is 
the inverse temperature, ${\bf v}({\bf k})$ denotes the Fermi velocity calculated 
from the charge self-consistent Bloch bands and $f_{\mu}$ marks the 
Fermi-Dirac distribution. To extract the $k$-resolved one-particle spectral function 
$A({\bf k},\omega)=-\pi^{-1}\,{\rm Im}\,G({\bf k},\omega)$ from the Green's
function $G$, an analytical continuation of the electronic self-energy in 
Matsubara space $\omega_n$ is performed via the Pad{\'e} approximation. 
For more details we refer to Ref.~\cite{boe14}. This framework allows us to 
compute the thermopower through 
$S=-\frac{{\rm k}^{\hfill}_{\rm B}}{|e|}\frac{K_1}{K_0}\;,$ and the resistivity 
as $\rho=\frac{1}{K_0}$, both beyond the constant-relaxation-time approximation.

\paragraph{Materials preparation, characterization and measurement}.
\begin{figure}[t]
\begin{center}
\hspace*{-0.2cm}\includegraphics*[width=7cm]{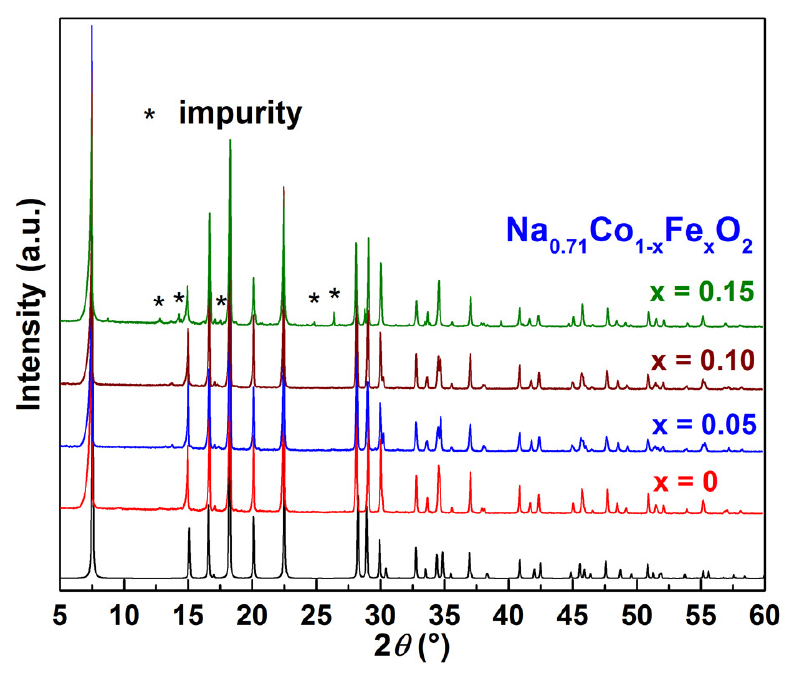}
\end{center}
\caption{(color online) XRD pattern of the 
Na$_{0.71}$Co$_{1-y}$Fe$_y$O$_2$ ($y=0, 0.05, 0.10$ and $0.15)$ powder samples. 
Black line: reference data obtained from the ICSD database.} 
\label{fig2:xrd}
\end{figure}
The powders of Na$_{0.71}$Co$_{1-y}$Fe$_y$O$_2$ ($y=0, 0.05, 0.10,$ and 0.15) are prepared by 
the Pechini method. Stoichiometric amounts of the chemicals: sodium acetate (NaC$_2$H$_3$O$_2$,  
99.0\%), Iron (III) nitrate nonahydrate (Sigma-Aldrich, purity  98\%), cobalt acetate tetrahydrate 
((CH$_3$COO)$_2$Co$\cdot$4 H$_2$O,  98.0\%) and citric acid (C$_6$H$_8$O$_7$, 99 \%, CA) are each 
dissolved in deionized distilled water. The calculated amount of citric acid and ingredient 
acetates is mandated at a molar ratio of 1.5:1. The precursor solution is heated in an oil bath 
at 65$^\circ$\,C while stirring continuously until a uniform viscous gel (2-3 h) forms. 
Subsequently, ethylene glycol (HOCH$_2$CH$_2$OH, EG, $\rho=1.11$ g/cm$^3$) is added  
as a gelling agent in a molar ratio of 3:1 to the amount of citric acid. The gel is dried 
at 120$^\circ$\,C for 4 hours and then heated in air to 300$^\circ$\,C for 10 hours. 
The resulting powder is calcinated in air at 850$^\circ$\,C for 20 hours. Densification was 
done in two steps: cold pressing in air (20 kN, 20 min) followed by iso-static pressing 
(800 kN, 1 min). Finally, the pellets are sintered for 24 hours at 900$^\circ$C in air. 

The samples' phase structure is identified by the Bruker D8-Advance diffractometer 
in Debye-Scherrer geometry with (220) Ge monochromator (Mo-K$\alpha$1, x-ray wavelength of 0.70930\AA). 
Seebeck coefficient and resistivity are measured simultaneously with a ZEM-3 (M10) 
ULVAC system, supplied with Pt electrodes, in the range $T=30^\circ-350^\circ$\,C and 950 mBar 
oxygen pressure. The uncertainties for both transport properties amount to $\pm7$\%.

XRD patterns are plotted in Fig.~\ref{fig2:xrd}. For Fe substitution $>10\%$ an impurity phase 
occurs. Thus the experimental study of of phase-pure Fe-doped samples is here limited to $y\le 0.1$. 
Note that an Fe doping of Na$_{0.63}$CoO$_2$ with $y\le 0.03$ has been reported by 
Zhou {\sl et al.}~\cite{zho10}, but without thermoelectric characterization.

\begin{figure}[b]
\includegraphics*[width=8cm]{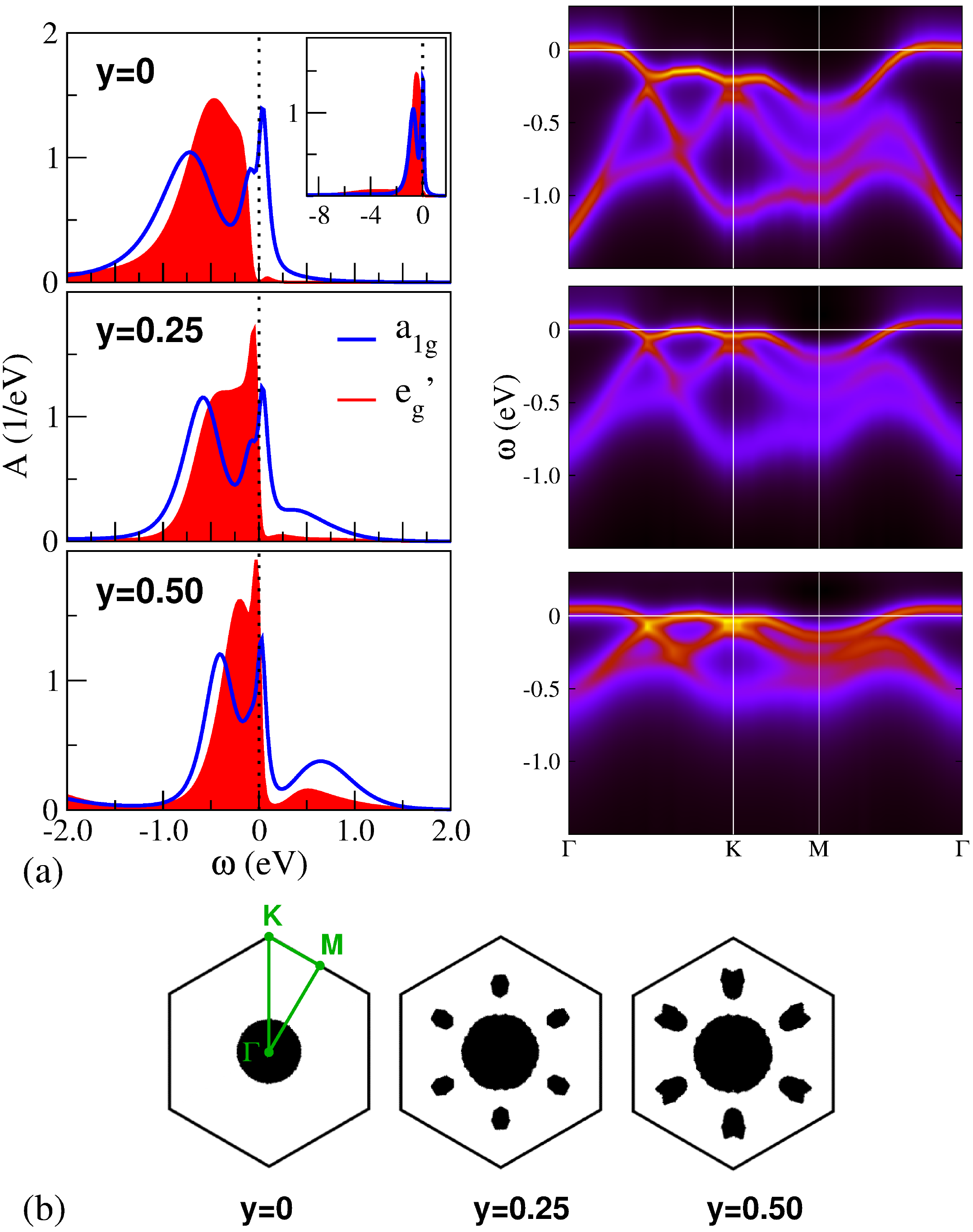}
\caption{(color online)
DFT+DMFT electronic structure for Na$_{0.7}$Co$_{1-y}$Fe$_{y}$O$_2$. 
(a) Spectral function for different $y$. 
Left: $k$-integrated $A$=$\sum_{\bf k}A({\bf k},\omega)$ and 
right:  $k$-resolved along high-symmetry lines in the Brillouin zone. 
(b) $y$-dependent FS.}\label{fig3:spectra}
\end{figure}

\paragraph{Correlated electronic structure.}
The Na$_{0.7}$CoO$_2$ electronic structure with multi-orbital DMFT
self-energy effects has been discussed in Ref.~\cite{boe14}. Electronic
correlations are effective in renormalizing the low-energy $t_{2g}$ band manifold, 
introducing broadening due to finite lifetime effects and shifting the occupied
$e_g$ pockets further away from the Fermi level (see top of Fig~\ref{fig3:spectra}a). 
The single-sheet hole-like FS stems from an $a_{1g}$ dominated band.

Replacing a fraction $y$ of cobalt by iron introduces hole doping of the same
concentration. The actual transition-metal $t_{2g}$ filling then reads
$n(t_{2g})=5+x-y\equiv 5+\tilde{x}$. We here employ two rather large theoretical Fe 
dopings of $y=0.25, 0.50$ to illustrate the principle effect. Because of the averaged
VCA treatment, we underestimate its realistic impact and thus the lower experimental 
dopings yield comparable effective physics. Figure~\ref{fig3:spectra} exhibits the changes
of the spectral function due to hole doping $y$.
The $e_g'$ pockets shift towards $\varepsilon_{\rm F}$, develop a low-energy quasiparticle 
(QP) resonance and participate already for $y$=0.25 in the FS. Since the partial 
bandwidth of the $e_g'$ derived states is reduced compared to the $a_{1g}$ one due to smaller 
hopping, the pocket QP resonance is rather sharp. Additionally the overall 
renormalization is enhanced with $y$ and an upper Hubbard band sets in for $y$=0.50. 
This strengthening of electronic correlations is not surprising since the Fe doping
effectively drives Na$_{0.7}$CoO$_2$ again further away from the band-insulating ($x$=1)
limit. However importantly, note that systems at a sole Na-doping level $x_1$ and at
effective doping level $\tilde{x}_1=x_2$$-$$y\stackrel{!}{=}x_1$ are {\sl not} 
electronically identical. 
Lets assume to lower the electron doping starting from $x$=0.7 where hole pockets are shifted
below $\varepsilon_{\rm F}$. Then it was shown in Ref.~\cite{boe14} that for $x$=0.3 the 
$e_g'$ pockets are still surpressed at the Fermi level. But here for 
$\tilde{x}$=$0.7-0.25$=0.45 these pockets already cross $\varepsilon_{\rm F}$. Thus Fe 
doping strengthens the multi-orbital transport character compared to pure-Na doping.  

\paragraph{Theoretical transport.}
\begin{figure}[t]
\includegraphics*[width=8.2cm]{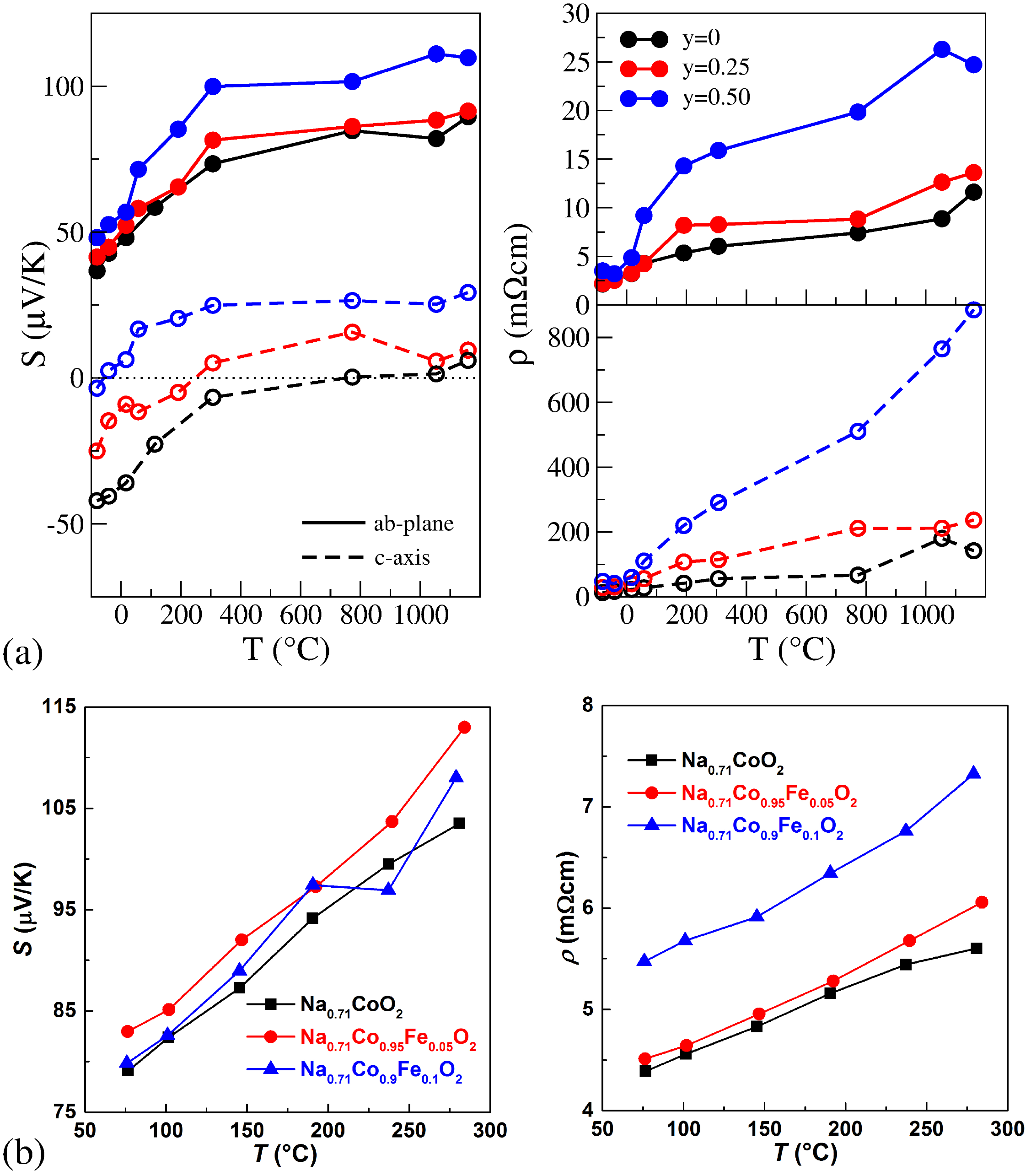}
\caption{(color online) Thermopower and resistivity for different Fe doping $y$.
(a) Theoretical data from DFT+DMFT for $y=0, 0.25, 0.50$. Solid lines: in-plane, 
dashed lines: along $c$-axis. 
(b) Experimental data for $y=0, 0.05, 0.10.$}\label{fig4:see-res}
\end{figure}
In Ref.~\cite{boe14} the anisotropic thermopower of sodium cobaltate at $x=0.70, 0.75$  
has been revealed using the present DFT+DMFT-based multi-orbital Kubo formalism. 
Here we compare in Fig.~\ref{fig4:see-res}a the results with and without Fe doping for 
Na$_{0.7}$CoO$_2$. Quantitatively, the in-plane Seebeck values without Fe doping match 
the experimental data by Kaurav {\sl et al.}~\cite{kau09}. As a proof of principles, theory 
documents an enhancement of the in-plane thermoelectric transport compared to the Fe-free 
case. For once, it may be explained by the increase of transport-relevant electron-hole 
asymmetry due to the emerging $e_g'$ pockets at $\varepsilon_{\rm F}$. A Seebeck increase 
because of  this effect was predicted early on by model studies~\cite{kur07}, though its 
realization by Fe doping was not forseen. In addition, the small negative $c$-axis thermopower 
in Fe-free Na$_{0.7}$CoO$_2$ eventually changes sign for larger $y$. This fosters
the coherency of the net thermoelectric transport.  

The resistivity also increases with Fe doping and the in-plane values
match our experimental data (see below). A gain of scattering because of
reinforced electron correlations and appearing interband processes is to be blamed. 
Though the $c$-axis resistivity largely exceeds the in-plane one, the transport anisotropy
 seems still underestimated compared to measurements by Wang {\sl et al.}~\cite{wan05}.

\paragraph{Experimental transport.}
In direct comparison to the theoretical data, Fig.~\ref{fig4:see-res}b displays the experimentally
determined Seebeck coefficient. A clear observation can be extracted, namely 
Fe-doping gives rise to an increase of the experimental thermopower compared to the Fe-free case. 
But this enhancement is not of monotonic form with $y$ and as already remarked, 
the achievable phase-pure Fe content is experimentally limited to $y$=0.1. 
Still, experiment verifies qualitatively the theoretical prediction on a similar quantitative level
in terms of relative thermopower growth. The non-monotonic character might result from intricate
real-space effects, e.g. a modification of the Co/Fe local-moment landscape, which
are beyond the VCA method that was employed in theory.

The resistivity increase reported in Fig.~\ref{fig4:see-res}b also verifies the theoretically 
obtained scattering enhancement. More pronounced correlation effects, the introduced 
interband scattering for finite $y$ and/or modified vibrational properties are at the origin of this 
behavior. Also here, a deeper theoretical analysis beyond VCA, with additional inclusion of the phonon 
degrees of freedom, is in order to single out the dominate scattering mechanism.

\paragraph{Conclusions.}
Theory and experiment agree in the enhancement of the sodium cobaltate thermopower iron doping. 
This agreement may serve as a proof of principles for further engineering protocols. 
Our calculations for Fe-doped Na$_{0.7}$CoO$_2$ show that effective hole doping takes place. 
This shifts the originally occupied $e_g'$-like pockets to the Fermi level. The increased 
electron-hole transport asymmetry sustains an enlarged Seebeck coefficient. Iron doping also
increases electronic correlations by driving the system further away from the band-insulating 
regime. Yet the doping is different from a nominal identical hole doping via reducing Na content. 
This is not of complete surprise since first, the Na atom is here an electron donor compared to the 
Fe atom. Second, Na ions are positioned in between the CoO$_2$ planes and their contribution to 
bonding and scattering is rather different from substitutional Fe within the planes.

It would be interesting to further analyze the engineering possibilities of the cobaltate 
thermopower. Strobel {\sl et al.}~\cite{str09} already performed experimental studies for doping 
with Fe-isovalent Ru and also observed a non-monotonic thermopower increase, yet with semiconducting 
properties. A recent theoretical study of Sb-doped Na$_x$CoO$_2$ predicted a decrease in 
thermopower~\cite{ass15}. 
Generally, correlated materials are much more sensitive to doping than weakly correlated systems 
because of induced transfer of spectral weight.  
Another aspect in this regard is the nano-structuring of thermoelectric compounds~\cite{hic93,kou10}. 
Research on such structurings for the case of correlated materials, combined with a tailored chemical 
doping, in view of increasing the figure of merit is still at its infancy. The novel field of 
oxide heterostructures could serve as an ideal playground.

\begin{acknowledgments}
We thank L. Boehnke, who coded the Kubo framework in our DFT+DMFT scheme, for
very helpful discussions. We are indebted to D. Grieger for further theory discussions,
P. Yordanov and H.-U. Habermeier for help with the transport measurements, 
as well as X. Xiao, J. Moreno Mendaza, and M. Wiedemeyer for assistance in the 
sample preparation.
Financial support from DFG-SPP1386 and WE 2803/2-2 is acknowledged. Computations 
were performed at the University of Hamburg and the JURECA Cluster of the J\"ulich
Supercomputing Centre (JSC) under project number hhh08.
\end{acknowledgments}

\bibliography{bibextra}

\begin{thebibliography}{44}
\expandafter\ifx\csname natexlab\endcsname\relax\def\natexlab#1{#1}\fi
\expandafter\ifx\csname bibnamefont\endcsname\relax
  \def\bibnamefont#1{#1}\fi
\expandafter\ifx\csname bibfnamefont\endcsname\relax
  \def\bibfnamefont#1{#1}\fi
\expandafter\ifx\csname citenamefont\endcsname\relax
  \def\citenamefont#1{#1}\fi
\expandafter\ifx\csname url\endcsname\relax
  \def\url#1{\texttt{#1}}\fi
\expandafter\ifx\csname urlprefix\endcsname\relax\def\urlprefix{URL }\fi
\providecommand{\bibinfo}[2]{#2}
\providecommand{\eprint}[2][]{\url{#2}}

\bibitem[{\citenamefont{Terasaki et~al.}(1997)\citenamefont{Terasaki, Sasago,
  and Uchinokura}}]{ter97}
\bibinfo{author}{\bibfnamefont{I.}~\bibnamefont{Terasaki}},
  \bibinfo{author}{\bibfnamefont{Y.}~\bibnamefont{Sasago}}, \bibnamefont{and}
  \bibinfo{author}{\bibfnamefont{K.}~\bibnamefont{Uchinokura}},
  \bibinfo{journal}{Phys. Rev. B} \textbf{\bibinfo{volume}{56}},
  \bibinfo{pages}{R12685} (\bibinfo{year}{1997}).

\bibitem[{\citenamefont{Motohashi et~al.}(2001)\citenamefont{Motohashi,
  Naujalis, Ueda et~al.}}]{mot01}
\bibinfo{author}{\bibfnamefont{T.}~\bibnamefont{Motohashi}},
  \bibinfo{author}{\bibfnamefont{E.}~\bibnamefont{Naujalis}},
  \bibinfo{author}{\bibfnamefont{R.}~\bibnamefont{Ueda}}, \bibnamefont{et~al.},
  \bibinfo{journal}{Appl. Phys. Lett.} \textbf{\bibinfo{volume}{79}},
  \bibinfo{pages}{1480} (\bibinfo{year}{2001}).

\bibitem[{\citenamefont{Kaurav et~al.}(2009)\citenamefont{Kaurav, Wu, Kuo
  et~al.}}]{kau09}
\bibinfo{author}{\bibfnamefont{N.}~\bibnamefont{Kaurav}},
  \bibinfo{author}{\bibfnamefont{K.~K.} \bibnamefont{Wu}},
  \bibinfo{author}{\bibfnamefont{Y.~K.} \bibnamefont{Kuo}},
  \bibnamefont{et~al.}, \bibinfo{journal}{Phys. Rev. B}
  \textbf{\bibinfo{volume}{79}}, \bibinfo{pages}{075105}
  (\bibinfo{year}{2009}).

\bibitem[{\citenamefont{Lee et~al.}(2006)\citenamefont{Lee, Viciu, Li
  et~al.}}]{lee06}
\bibinfo{author}{\bibfnamefont{M.}~\bibnamefont{Lee}},
  \bibinfo{author}{\bibfnamefont{L.}~\bibnamefont{Viciu}},
  \bibinfo{author}{\bibfnamefont{L.}~\bibnamefont{Li}}, \bibnamefont{et~al.},
  \bibinfo{journal}{Nat. Mat.} \textbf{\bibinfo{volume}{5}},
  \bibinfo{pages}{537} (\bibinfo{year}{2006}).

\bibitem[{\citenamefont{H{\'e}bert et~al.}(2013)\citenamefont{H{\'e}bert,
  Kobayashi, Muguerra et~al.}}]{heb13}
\bibinfo{author}{\bibfnamefont{S.}~\bibnamefont{H{\'e}bert}},
  \bibinfo{author}{\bibfnamefont{W.}~\bibnamefont{Kobayashi}},
  \bibinfo{author}{\bibfnamefont{H.}~\bibnamefont{Muguerra}},
  \bibnamefont{et~al.}, \bibinfo{journal}{Phys. Status Solidi A}
  \textbf{\bibinfo{volume}{210}}, \bibinfo{pages}{69} (\bibinfo{year}{2013}).

\bibitem[{\citenamefont{Mukhamedshin et~al.}(2005)\citenamefont{Mukhamedshin,
  Alloul, Collin et~al.}}]{muk05}
\bibinfo{author}{\bibfnamefont{I.~R.} \bibnamefont{Mukhamedshin}},
  \bibinfo{author}{\bibfnamefont{H.}~\bibnamefont{Alloul}},
  \bibinfo{author}{\bibfnamefont{G.}~\bibnamefont{Collin}},
  \bibnamefont{et~al.}, \bibinfo{journal}{Phys. Rev. Lett.}
  \textbf{\bibinfo{volume}{94}}, \bibinfo{pages}{247602}
  (\bibinfo{year}{2005}).

\bibitem[{\citenamefont{Lang et~al.}(2008)\citenamefont{Lang, Bobroff, Alloul
  et~al.}}]{lan08}
\bibinfo{author}{\bibfnamefont{G.}~\bibnamefont{Lang}},
  \bibinfo{author}{\bibfnamefont{J.}~\bibnamefont{Bobroff}},
  \bibinfo{author}{\bibfnamefont{H.}~\bibnamefont{Alloul}},
  \bibnamefont{et~al.}, \bibinfo{journal}{Phys. Rev. B}
  \textbf{\bibinfo{volume}{78}}, \bibinfo{pages}{155116}
  (\bibinfo{year}{2008}).

\bibitem[{\citenamefont{Foo et~al.}(2004)\citenamefont{Foo, Wang, Watauchi
  et~al.}}]{foo04}
\bibinfo{author}{\bibfnamefont{M.~L.} \bibnamefont{Foo}},
  \bibinfo{author}{\bibfnamefont{Y.}~\bibnamefont{Wang}},
  \bibinfo{author}{\bibfnamefont{S.}~\bibnamefont{Watauchi}},
  \bibnamefont{et~al.}, \bibinfo{journal}{Phys. Rev. Lett.}
  \textbf{\bibinfo{volume}{92}}, \bibinfo{pages}{247001}
  (\bibinfo{year}{2004}).

\bibitem[{\citenamefont{Motohashi et~al.}(2003)\citenamefont{Motohashi, Ueda,
  Naujalis et~al.}}]{mot03}
\bibinfo{author}{\bibfnamefont{T.}~\bibnamefont{Motohashi}},
  \bibinfo{author}{\bibfnamefont{R.}~\bibnamefont{Ueda}},
  \bibinfo{author}{\bibfnamefont{E.}~\bibnamefont{Naujalis}},
  \bibnamefont{et~al.}, \bibinfo{journal}{Phys. Rev. B}
  \textbf{\bibinfo{volume}{67}}, \bibinfo{pages}{064406}
  (\bibinfo{year}{2003}).

\bibitem[{\citenamefont{Boothroyd et~al.}(2004)\citenamefont{Boothroyd, Coldea,
  Tennant et~al.}}]{boo04}
\bibinfo{author}{\bibfnamefont{A.~T.} \bibnamefont{Boothroyd}},
  \bibinfo{author}{\bibfnamefont{R.}~\bibnamefont{Coldea}},
  \bibinfo{author}{\bibfnamefont{D.~A.} \bibnamefont{Tennant}},
  \bibnamefont{et~al.}, \bibinfo{journal}{Phys. Rev. Lett.}
  \textbf{\bibinfo{volume}{92}}, \bibinfo{pages}{197201}
  (\bibinfo{year}{2004}).

\bibitem[{\citenamefont{Sakurai et~al.}(2004)\citenamefont{Sakurai, Tsujii, and
  Takayama-Muromachi}}]{sak04}
\bibinfo{author}{\bibfnamefont{H.}~\bibnamefont{Sakurai}},
  \bibinfo{author}{\bibfnamefont{N.}~\bibnamefont{Tsujii}}, \bibnamefont{and}
  \bibinfo{author}{\bibfnamefont{E.}~\bibnamefont{Takayama-Muromachi}},
  \bibinfo{journal}{J. Phys. Soc. Jpn.} \textbf{\bibinfo{volume}{73}},
  \bibinfo{pages}{2393} (\bibinfo{year}{2004}).

\bibitem[{\citenamefont{Bayrakci et~al.}(2005)\citenamefont{Bayrakci, Mirebeau,
  Bourges et~al.}}]{bay05}
\bibinfo{author}{\bibfnamefont{S.~P.} \bibnamefont{Bayrakci}},
  \bibinfo{author}{\bibfnamefont{I.}~\bibnamefont{Mirebeau}},
  \bibinfo{author}{\bibfnamefont{P.}~\bibnamefont{Bourges}},
  \bibnamefont{et~al.}, \bibinfo{journal}{Phys. Rev. Lett.}
  \textbf{\bibinfo{volume}{94}}, \bibinfo{pages}{157205}
  (\bibinfo{year}{2005}).

\bibitem[{\citenamefont{Helme et~al.}(2006)\citenamefont{Helme, Boothroyd,
  Coldea et~al.}}]{hel06}
\bibinfo{author}{\bibfnamefont{L.~M.} \bibnamefont{Helme}},
  \bibinfo{author}{\bibfnamefont{A.~T.} \bibnamefont{Boothroyd}},
  \bibinfo{author}{\bibfnamefont{R.}~\bibnamefont{Coldea}},
  \bibnamefont{et~al.}, \bibinfo{journal}{Phys. Rev. B}
  \textbf{\bibinfo{volume}{73}}, \bibinfo{pages}{054405}
  (\bibinfo{year}{2006}).

\bibitem[{\citenamefont{Shu et~al.}(2007)\citenamefont{Shu, Prodi, Chu
  et~al.}}]{shu07}
\bibinfo{author}{\bibfnamefont{G.~J.} \bibnamefont{Shu}},
  \bibinfo{author}{\bibfnamefont{A.}~\bibnamefont{Prodi}},
  \bibinfo{author}{\bibfnamefont{S.~Y.} \bibnamefont{Chu}},
  \bibnamefont{et~al.}, \bibinfo{journal}{Phys. Rev. B}
  \textbf{\bibinfo{volume}{76}}, \bibinfo{pages}{184115}
  (\bibinfo{year}{2007}).

\bibitem[{\citenamefont{Schulze et~al.}(2008)\citenamefont{Schulze,
  Br{\"u}hwiler, H{\"a}fliger et~al.}}]{schu08}
\bibinfo{author}{\bibfnamefont{T.~F.} \bibnamefont{Schulze}},
  \bibinfo{author}{\bibfnamefont{M.}~\bibnamefont{Br{\"u}hwiler}},
  \bibinfo{author}{\bibfnamefont{P.~S.} \bibnamefont{H{\"a}fliger}},
  \bibnamefont{et~al.}, \bibinfo{journal}{Phys. Rev. B}
  \textbf{\bibinfo{volume}{78}}, \bibinfo{pages}{205101}
  (\bibinfo{year}{2008}).

\bibitem[{\citenamefont{Wilhelm et~al.}(2015)\citenamefont{Wilhelm, Lechermann,
  Hafermann et~al.}}]{wil15}
\bibinfo{author}{\bibfnamefont{A.}~\bibnamefont{Wilhelm}},
  \bibinfo{author}{\bibfnamefont{F.}~\bibnamefont{Lechermann}},
  \bibinfo{author}{\bibfnamefont{H.}~\bibnamefont{Hafermann}},
  \bibnamefont{et~al.}, \bibinfo{journal}{Phys. Rev. B}
  \textbf{\bibinfo{volume}{91}}, \bibinfo{pages}{155114}
  (\bibinfo{year}{2015}).

\bibitem[{\citenamefont{Koshibae and Maekawa}(2001)}]{kos01}
\bibinfo{author}{\bibfnamefont{W.}~\bibnamefont{Koshibae}} \bibnamefont{and}
  \bibinfo{author}{\bibfnamefont{S.}~\bibnamefont{Maekawa}},
  \bibinfo{journal}{Phys. Rev. Lett.} \textbf{\bibinfo{volume}{87}},
  \bibinfo{pages}{236601} (\bibinfo{year}{2001}).

\bibitem[{\citenamefont{Xiang and Singh}(2007)}]{xia07}
\bibinfo{author}{\bibfnamefont{H.~J.} \bibnamefont{Xiang}} \bibnamefont{and}
  \bibinfo{author}{\bibfnamefont{D.~J.} \bibnamefont{Singh}},
  \bibinfo{journal}{Phys. Rev. B} \textbf{\bibinfo{volume}{76}},
  \bibinfo{pages}{195111} (\bibinfo{year}{2007}).

\bibitem[{\citenamefont{Hamada et~al.}(2007)\citenamefont{Hamada, Imai, and
  Funashima}}]{ham07}
\bibinfo{author}{\bibfnamefont{N.}~\bibnamefont{Hamada}},
  \bibinfo{author}{\bibfnamefont{T.}~\bibnamefont{Imai}}, \bibnamefont{and}
  \bibinfo{author}{\bibfnamefont{H.}~\bibnamefont{Funashima}},
  \textbf{\bibinfo{volume}{19}}, \bibinfo{pages}{365221}
  (\bibinfo{year}{2007}).

\bibitem[{\citenamefont{Kuroki and Arita}(2007)}]{kur07}
\bibinfo{author}{\bibfnamefont{K.}~\bibnamefont{Kuroki}} \bibnamefont{and}
  \bibinfo{author}{\bibfnamefont{R.}~\bibnamefont{Arita}}, \bibinfo{journal}{J.
  Phys. Soc. Jpn.} \textbf{\bibinfo{volume}{76}}, \bibinfo{pages}{083707}
  (\bibinfo{year}{2007}).

\bibitem[{\citenamefont{Peterson et~al.}(2007)\citenamefont{Peterson, Shastry,
  and Haerter}}]{pet07}
\bibinfo{author}{\bibfnamefont{M.~R.} \bibnamefont{Peterson}},
  \bibinfo{author}{\bibfnamefont{B.~S.} \bibnamefont{Shastry}},
  \bibnamefont{and} \bibinfo{author}{\bibfnamefont{J.~O.}
  \bibnamefont{Haerter}}, \bibinfo{journal}{Phys. Rev. B}
  \textbf{\bibinfo{volume}{76}}, \bibinfo{pages}{165118}
  (\bibinfo{year}{2007}).

\bibitem[{\citenamefont{Wissgott et~al.}(2010)\citenamefont{Wissgott, Toschi,
  Usui et~al.}}]{wis10}
\bibinfo{author}{\bibfnamefont{P.}~\bibnamefont{Wissgott}},
  \bibinfo{author}{\bibfnamefont{A.}~\bibnamefont{Toschi}},
  \bibinfo{author}{\bibfnamefont{H.}~\bibnamefont{Usui}}, \bibnamefont{et~al.},
  \bibinfo{journal}{Phys. Rev. B} \textbf{\bibinfo{volume}{82}},
  \bibinfo{pages}{201106(R)} (\bibinfo{year}{2010}).

\bibitem[{\citenamefont{Sangiovanni et~al.}(2012)\citenamefont{Sangiovanni,
  Wissgott, Assaad et~al.}}]{san12}
\bibinfo{author}{\bibfnamefont{G.}~\bibnamefont{Sangiovanni}},
  \bibinfo{author}{\bibfnamefont{P.}~\bibnamefont{Wissgott}},
  \bibinfo{author}{\bibfnamefont{F.}~\bibnamefont{Assaad}},
  \bibnamefont{et~al.}, \bibinfo{journal}{Phys. Rev. B}
  \textbf{\bibinfo{volume}{86}}, \bibinfo{pages}{035123}
  (\bibinfo{year}{2012}).

\bibitem[{\citenamefont{Hasan et~al.}(2004)\citenamefont{Hasan, Chuang, Qian
  et~al.}}]{has04}
\bibinfo{author}{\bibfnamefont{M.~Z.} \bibnamefont{Hasan}},
  \bibinfo{author}{\bibfnamefont{Y.-D.} \bibnamefont{Chuang}},
  \bibinfo{author}{\bibfnamefont{D.}~\bibnamefont{Qian}}, \bibnamefont{et~al.},
  \bibinfo{journal}{Phys. Rev. Lett.} \textbf{\bibinfo{volume}{92}},
  \bibinfo{pages}{246402} (\bibinfo{year}{2004}).

\bibitem[{\citenamefont{Geck et~al.}(2007)\citenamefont{Geck, Borisenko, Berger
  et~al.}}]{gec07}
\bibinfo{author}{\bibfnamefont{J.}~\bibnamefont{Geck}},
  \bibinfo{author}{\bibfnamefont{S.~V.} \bibnamefont{Borisenko}},
  \bibinfo{author}{\bibfnamefont{H.}~\bibnamefont{Berger}},
  \bibnamefont{et~al.}, \bibinfo{journal}{Phys. Rev. Lett.}
  \textbf{\bibinfo{volume}{99}}, \bibinfo{pages}{046403}
  (\bibinfo{year}{2007}).

\bibitem[{\citenamefont{Wang et~al.}(2008)\citenamefont{Wang, Dai, and
  Fang}}]{wan08}
\bibinfo{author}{\bibfnamefont{G.-T.} \bibnamefont{Wang}},
  \bibinfo{author}{\bibfnamefont{X.}~\bibnamefont{Dai}}, \bibnamefont{and}
  \bibinfo{author}{\bibfnamefont{Z.}~\bibnamefont{Fang}},
  \bibinfo{journal}{Phys. Rev. Lett.} \textbf{\bibinfo{volume}{101}},
  \bibinfo{pages}{066403} (\bibinfo{year}{2008}).

\bibitem[{\citenamefont{Boehnke and Lechermann}(2014)}]{boe14}
\bibinfo{author}{\bibfnamefont{L.}~\bibnamefont{Boehnke}} \bibnamefont{and}
  \bibinfo{author}{\bibfnamefont{F.}~\bibnamefont{Lechermann}},
  \bibinfo{journal}{phys. stat. sol. (a)} \textbf{\bibinfo{volume}{211}},
  \bibinfo{pages}{1267} (\bibinfo{year}{2014}).

\bibitem[{\citenamefont{Grieger et~al.}(2012)\citenamefont{Grieger, Piefke,
  Peil et~al.}}]{gri12}
\bibinfo{author}{\bibfnamefont{D.}~\bibnamefont{Grieger}},
  \bibinfo{author}{\bibfnamefont{C.}~\bibnamefont{Piefke}},
  \bibinfo{author}{\bibfnamefont{O.~E.} \bibnamefont{Peil}},
  \bibnamefont{et~al.}, \bibinfo{journal}{Phys. Rev. B}
  \textbf{\bibinfo{volume}{86}}, \bibinfo{pages}{155121}
  (\bibinfo{year}{2012}).

\bibitem[{\citenamefont{Louie et~al.}(1979)\citenamefont{Louie, Ho, and
  Cohen}}]{lou79}
\bibinfo{author}{\bibfnamefont{S.~G.} \bibnamefont{Louie}},
  \bibinfo{author}{\bibfnamefont{K.~M.} \bibnamefont{Ho}}, \bibnamefont{and}
  \bibinfo{author}{\bibfnamefont{M.~L.} \bibnamefont{Cohen}},
  \bibinfo{journal}{Phys. Rev. B} \textbf{\bibinfo{volume}{19}},
  \bibinfo{pages}{1774} (\bibinfo{year}{1979}).

\bibitem[{\citenamefont{Rubtsov et~al.}(2005)\citenamefont{Rubtsov, Savkin, and
  Lichtenstein}}]{rub05}
\bibinfo{author}{\bibfnamefont{A.~N.} \bibnamefont{Rubtsov}},
  \bibinfo{author}{\bibfnamefont{V.~V.} \bibnamefont{Savkin}},
  \bibnamefont{and} \bibinfo{author}{\bibfnamefont{A.~I.}
  \bibnamefont{Lichtenstein}}, \bibinfo{journal}{Phys. Rev. B}
  \textbf{\bibinfo{volume}{72}}, \bibinfo{pages}{035122}
  (\bibinfo{year}{2005}).

\bibitem[{\citenamefont{Werner et~al.}(2006)\citenamefont{Werner, Comanac, de'
  Medici et~al.}}]{wer06}
\bibinfo{author}{\bibfnamefont{P.}~\bibnamefont{Werner}},
  \bibinfo{author}{\bibfnamefont{A.}~\bibnamefont{Comanac}},
  \bibinfo{author}{\bibfnamefont{L.}~\bibnamefont{de' Medici}},
  \bibnamefont{et~al.}, \bibinfo{journal}{Phys. Rev. Lett.}
  \textbf{\bibinfo{volume}{97}}, \bibinfo{pages}{076405}
  (\bibinfo{year}{2006}).

\bibitem[{\citenamefont{Boehnke et~al.}(2011)\citenamefont{Boehnke, Hafermann,
  Ferrero et~al.}}]{boe11}
\bibinfo{author}{\bibfnamefont{L.}~\bibnamefont{Boehnke}},
  \bibinfo{author}{\bibfnamefont{H.}~\bibnamefont{Hafermann}},
  \bibinfo{author}{\bibfnamefont{M.}~\bibnamefont{Ferrero}},
  \bibnamefont{et~al.}, \bibinfo{journal}{Phys. Rev. B}
  \textbf{\bibinfo{volume}{84}}, \bibinfo{pages}{075145}
  (\bibinfo{year}{2011}).

\bibitem[{\citenamefont{Parcollet et~al.}(2015)\citenamefont{Parcollet,
  Ferrero, Ayral et~al.}}]{par15}
\bibinfo{author}{\bibfnamefont{O.}~\bibnamefont{Parcollet}},
  \bibinfo{author}{\bibfnamefont{M.}~\bibnamefont{Ferrero}},
  \bibinfo{author}{\bibfnamefont{T.}~\bibnamefont{Ayral}},
  \bibnamefont{et~al.}, \bibinfo{journal}{Comput. Phys. Commun.}
  \textbf{\bibinfo{volume}{196}}, \bibinfo{pages}{398} (\bibinfo{year}{2015}).

\bibitem[{\citenamefont{Amadon et~al.}(2008)\citenamefont{Amadon, Lechermann,
  Georges et~al.}}]{ama08}
\bibinfo{author}{\bibfnamefont{B.}~\bibnamefont{Amadon}},
  \bibinfo{author}{\bibfnamefont{F.}~\bibnamefont{Lechermann}},
  \bibinfo{author}{\bibfnamefont{A.}~\bibnamefont{Georges}},
  \bibnamefont{et~al.}, \bibinfo{journal}{Phys. Rev. B}
  \textbf{\bibinfo{volume}{77}}, \bibinfo{pages}{205112}
  (\bibinfo{year}{2008}).

\bibitem[{\citenamefont{Anisimov et~al.}(2005)\citenamefont{Anisimov, Kondakov,
  Kozhevnikov et~al.}}]{ani05}
\bibinfo{author}{\bibfnamefont{V.~I.} \bibnamefont{Anisimov}},
  \bibinfo{author}{\bibfnamefont{D.~E.} \bibnamefont{Kondakov}},
  \bibinfo{author}{\bibfnamefont{A.~V.} \bibnamefont{Kozhevnikov}},
  \bibnamefont{et~al.}, \bibinfo{journal}{Phys. Rev. B}
  \textbf{\bibinfo{volume}{71}}, \bibinfo{pages}{125119}
  (\bibinfo{year}{2005}).

\bibitem[{\citenamefont{Anisimov et~al.}(1993)\citenamefont{Anisimov, Solovyev,
  Korotin, Czy$\dot{\text{z}}$yk, and Sawatzky}}]{ani93}
\bibinfo{author}{\bibfnamefont{V.~I.} \bibnamefont{Anisimov}},
  \bibinfo{author}{\bibfnamefont{I.~V.} \bibnamefont{Solovyev}},
  \bibinfo{author}{\bibfnamefont{M.~A.} \bibnamefont{Korotin}},
  \bibinfo{author}{\bibfnamefont{M.~T.} \bibnamefont{Czy$\dot{\text{z}}$yk}},
  \bibnamefont{and} \bibinfo{author}{\bibfnamefont{G.~A.}
  \bibnamefont{Sawatzky}}, \bibinfo{journal}{Phys. Rev. B}
  \textbf{\bibinfo{volume}{48}}, \bibinfo{pages}{16929} (\bibinfo{year}{1993}).

\bibitem[{\citenamefont{Oudovenko et~al.}(2006)\citenamefont{Oudovenko,
  P{\'a}lsson, Haule et~al.}}]{oud06}
\bibinfo{author}{\bibfnamefont{V.~S.} \bibnamefont{Oudovenko}},
  \bibinfo{author}{\bibfnamefont{G.}~\bibnamefont{P{\'a}lsson}},
  \bibinfo{author}{\bibfnamefont{K.}~\bibnamefont{Haule}},
  \bibnamefont{et~al.}, \bibinfo{journal}{Phys. Rev. B}
  \textbf{\bibinfo{volume}{73}}, \bibinfo{pages}{035120}
  (\bibinfo{year}{2006}).

\bibitem[{\citenamefont{Deng et~al.}(2013)\citenamefont{Deng, Mravlje, {\v
  Z}itko et~al.}}]{den13}
\bibinfo{author}{\bibfnamefont{X.}~\bibnamefont{Deng}},
  \bibinfo{author}{\bibfnamefont{J.}~\bibnamefont{Mravlje}},
  \bibinfo{author}{\bibfnamefont{R.}~\bibnamefont{{\v Z}itko}},
  \bibnamefont{et~al.}, \bibinfo{journal}{Phys. Rev. Lett.}
  \textbf{\bibinfo{volume}{110}}, \bibinfo{pages}{086401}
  (\bibinfo{year}{2013}).

\bibitem[{\citenamefont{Zhou et~al.}(2010)\citenamefont{Zhou, Zhang, Button
  et~al.}}]{zho10}
\bibinfo{author}{\bibfnamefont{T.}~\bibnamefont{Zhou}},
  \bibinfo{author}{\bibfnamefont{D.}~\bibnamefont{Zhang}},
  \bibinfo{author}{\bibfnamefont{T.~W.} \bibnamefont{Button}},
  \bibnamefont{et~al.}, \bibinfo{journal}{Dalton Trans.}
  \textbf{\bibinfo{volume}{39}}, \bibinfo{pages}{1089} (\bibinfo{year}{2010}).

\bibitem[{\citenamefont{Wang et~al.}(2005)\citenamefont{Wang, Chen, Luo
  et~al.}}]{wan05}
\bibinfo{author}{\bibfnamefont{C.~H.} \bibnamefont{Wang}},
  \bibinfo{author}{\bibfnamefont{X.~H.} \bibnamefont{Chen}},
  \bibinfo{author}{\bibfnamefont{J.~L.} \bibnamefont{Luo}},
  \bibnamefont{et~al.}, \bibinfo{journal}{Phys. Rev. B}
  \textbf{\bibinfo{volume}{71}}, \bibinfo{pages}{224515}
  (\bibinfo{year}{2005}).

\bibitem[{\citenamefont{Strobel et~al.}(2009)\citenamefont{Strobel, Muguerra,
  H{\'e}bert et~al.}}]{str09}
\bibinfo{author}{\bibfnamefont{P.}~\bibnamefont{Strobel}},
  \bibinfo{author}{\bibfnamefont{H.}~\bibnamefont{Muguerra}},
  \bibinfo{author}{\bibfnamefont{S.}~\bibnamefont{H{\'e}bert}},
  \bibnamefont{et~al.}, \bibinfo{journal}{J. of Solid State Chem.}
  \textbf{\bibinfo{volume}{182}}, \bibinfo{pages}{1872} (\bibinfo{year}{2009}).

\bibitem[{\citenamefont{Assadi and Katayama-Yoshida}(2015)}]{ass15}
\bibinfo{author}{\bibfnamefont{M.~H.~N.} \bibnamefont{Assadi}}
  \bibnamefont{and}
  \bibinfo{author}{\bibfnamefont{H.}~\bibnamefont{Katayama-Yoshida}},
  \bibinfo{journal}{J. Phys.: Condens. Matter} \textbf{\bibinfo{volume}{27}},
  \bibinfo{pages}{175504} (\bibinfo{year}{2015}).

\bibitem[{\citenamefont{Hicks and Dresselhaus}(1993)}]{hic93}
\bibinfo{author}{\bibfnamefont{L.~D.} \bibnamefont{Hicks}} \bibnamefont{and}
  \bibinfo{author}{\bibfnamefont{M.~S.} \bibnamefont{Dresselhaus}},
  \bibinfo{journal}{Phys. Rev. B} \textbf{\bibinfo{volume}{47}},
  \bibinfo{pages}{12727} (\bibinfo{year}{1993}).

\bibitem[{\citenamefont{Koumoto et~al.}(2010)\citenamefont{Koumoto, Wang, Zhang
  et~al.}}]{kou10}
\bibinfo{author}{\bibfnamefont{K.}~\bibnamefont{Koumoto}},
  \bibinfo{author}{\bibfnamefont{Y.}~\bibnamefont{Wang}},
  \bibinfo{author}{\bibfnamefont{R.}~\bibnamefont{Zhang}},
  \bibnamefont{et~al.}, \bibinfo{journal}{Annu. Rev. Mater. Res.}
  \textbf{\bibinfo{volume}{40}}, \bibinfo{pages}{363} (\bibinfo{year}{2010}).

\end{thebibliography}

\end{document}